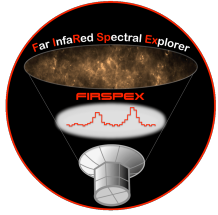

# The Far-InfraRed Spectroscopic Explorer (FIRSPEX)

**Dimitra Rigopoulou[1,2], Chris Pearson[1,2], Brian Ellison[2], Bruce Swinyard[2,3], Sheng-Cai Shi[4], Jie Hu[4], Martin Caldwell[2], Jia-Sheng Huang[5] and the FIRSPEX Consortium[6]**

[1]Astrophysics, University of Oxford, UK,
[2]Rutherford Appleton Laboratory, UK,
[3]University College London, UK
[4]Purple Mountain Observatory, CAS, China,
[5]National Astronomical Observatories of China, CAS, China
[6] includes scientists from the UK, China, France, Greece, Italy, & the Netherlands

## Abstract
The Far-InfraRed Spectroscopic Explorer (FIRSPEX) is a candidate mission in response to a bi-lateral Small-mission call issued by the European Space Agency (ESA) and the Chinese Academy of Sciences (CAS). FIRSPEX is a small satellite (~1m telescope) operating from Low Earth Orbit (LEO). It consists of a number of heterodyne detection bands targeting key molecular and atomic transitions in the terahertz (THz) and Supra-Terahertz (>1 THz) frequency range. The FIRSPEX bands are: [CII] 158 microns (1.9 THz), [NII] 205 microns (1.46 THz), [CI] 370 microns (0.89 THz), CO(6-5) 433 microns (0.69 THz). The primary goal of FIRSPEX is to perform an `unbiased' all sky spectroscopic survey in four far-infrared lines delivering the first 3D-maps (high spectral resolution) of the Galaxy. The spectroscopic surveys will build on the heritage of Herschel and complement the broad-band all-sky surveys carried out by the IRAS and AKARI observatories. In addition FIRSPEX will enable targeted observations of nearby and distant galaxies allowing for an in-depth study of the ISM components.

## 1.     Background

The far-infrared (FIR) to submillimetre (submm) window is one of the least-studied regions of the electromagnetic spectrum. Yet, this wavelength range is absolutely crucial to our understanding of star formation and stellar evolution in the Universe. These complex physical processes leave their imprint on the Interstellar medium (ISM) of the Milky Way and external galaxies. By studying the phase structure of the ISM we can begin to unravel the processes that control the heating and cooling of the clouds that regulate star formation. FIR/submm spectroscopy is an invaluable tool in this study since this regime contains important cooling lines of the different phases of the ISM.

The fine-structure line of singly ionised carbon [CII] at 157.7μm is the most important cooling line of the neutral ISM and is the brightest line in the spectrum of the Milky Way and external galaxies. FIRAS/COBE maps showed that this is the strongest cooling line in the ISM at about 0.3% of the continuum infrared emission. Since ionized carbon can be found throughout the ISM and can be excited by collisions with electrons, atomic hydrogen (HI), and neutral hydrogen (H) the [CII] line traces the warm ionized medium, the warm and cold diffuse atomic medium and warm and dense molecular gas.

The far-IR fine structure line of ionized nitrogen (N+ 205 μm) is a coolant of the diffuse ionized medium. But N+ can also be used to determine the amount of C+ that comes from ionized clouds. The N+ and the C+ lines have nearly identical critical densities for excitation in ionized gas regions. Their line ratio is thus insensitive to the hardness of the stellar radiation field (since the photon energies required to ionize each species to the next ionization states are similar) and is only a function of the N+/C+ abundance ratio. Therefore, N+ can be used to measure the fraction of the observed C+ that arises from ionized gas, thereby better describing the cooling of both the ionized gas and the neutral ISM. Neutral carbon (C 370 μm) originates in the transition layers in the surfaces of molecular clouds – including photo-dissociation regions – and probes the transformation of C+ to carbon monoxide (CO). The CO (middle level, hereafter mid-J) transitions probe relatively dense and warm gas usually found in regions illuminated by massive stars. The mid- to high-J transitions, which are either difficult to observe or completely inaccessible from the ground, have a large span in critical densities, making them excellent tracers of the physical conditions of gas over a wide range in temperatures and densities.

These far-infrared lines are key to understanding the complex physics of the gas phases of the ISM. In recent years, there has been an effort to explore the properties of the ISM in a variety of environments from the Galaxy to nearby and distant external galaxies. And while the recent advent of Herschel (Pilbratt et al. 2010) combined with the availability of the Atacama Large Millimetre Array (ALMA) has enabled observations of some of these lines and subsequent analysis, a systematic wide area survey aimed at large scale emission and distribution of the gas composition of the ISM is still to come. The **F**ar-**I**nfra**R**ed **Sp**ectroscoic **Ex**plorer (FIRSPEX) concept aims to fill this gap by providing key insight in the life-cycle of interstellar gas in the Milky Way and nearby external galaxies.

## 2. From COBE to FIRSPEX

The FIRAS instrument on the COBE satellite (Boggess et al. 1992) spectroscopically surveyed the entire sky from 0.1 to 10mm with a spatial resolution of 7º and a velocity resolution of 1000 km/s (Figure 1). FIRAS determined that the C+ line is the dominant cooling line of the ISM at ~0.3% of the continuum infrared emission while the N+ & C lines are less intense by a factor of 10 and 100, respectively. As reported by Fixsen, Bennett, & Mather (1999), all three lines peak in the central regions of our galaxy (the molecular ring) and originate from a layer with a thickness of a few degrees but the mission had insufficient resolution to relate the emission to specific components and sources within the ISM. The Balloon-borne Infrared Carbon Explorer (BICE, Nakagawa et al. 1998) mapped some 100 square degrees around the Galactic Centre with a spatial resolution of 15' and a velocity resolution of 175 km/s. The BICE results revealed that the diffuse ISM is more emissive in C+ relative to the far-IR dust emission than compact regions of massive star formation or the Galactic Centre, further emphasizing the importance of the cold neutral medium for the total C+ emission from the Milky Way. FIRAS/COBE was instrumental in determining the global characteristics of these atomic line tracers and brought the key questions that FIRSPEX will address to the forefront of ISM studies.

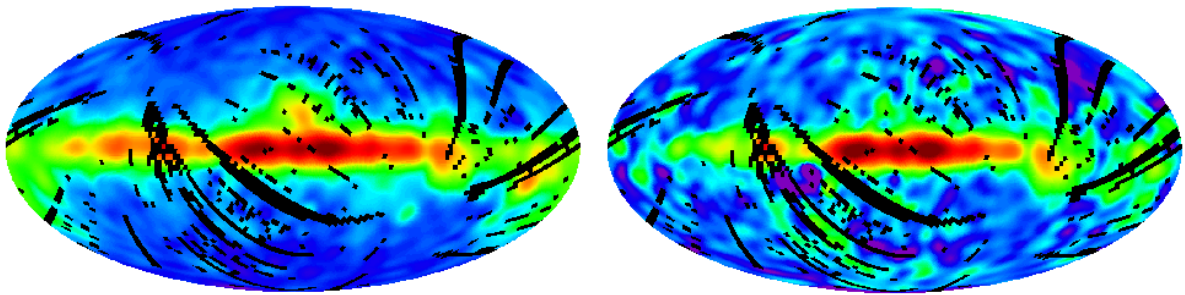

**Figure 1:** COBE/FIRAS maps of the C+ (left) and N+ (right) line intensities. The low velocity resolution did not allow velocities from individual sources to be measured (Fixsen, Bennett and Mather 1999).

The availability of sub–kms$^{-1}$ resolution with the heterodyne instrument HIFI (de Graauw et al. 2010) on Herschel, together with the 12″ angular resolution afforded by its 3.5m diameter telescope, enabled the first spatially resolved, high spectral resolution observations of C+ towards various LOS in the Galaxy. In contrast to the global surveys, this very sparse survey emphasized the importance of C+ emission in these sightlines from spiral arms as well as from regions of massive star formation. The Galactic Observations of the Terahertz-C+ (GOTC+) project (e.g. Langer 2010, Pineda 2010, Velusamy 2012) targeted some 450 LOS covering 360° in longitude, but not targeting specific regions and, of course, was highly under-sampled in angle (Figure 2). Mostly, though, with its high spectral resolution which allowed identification of individual clouds, this survey provided a first view of the tremendous advances in our understanding of the physics of the ISM that awaits a full C+ survey of the Galactic Plane.

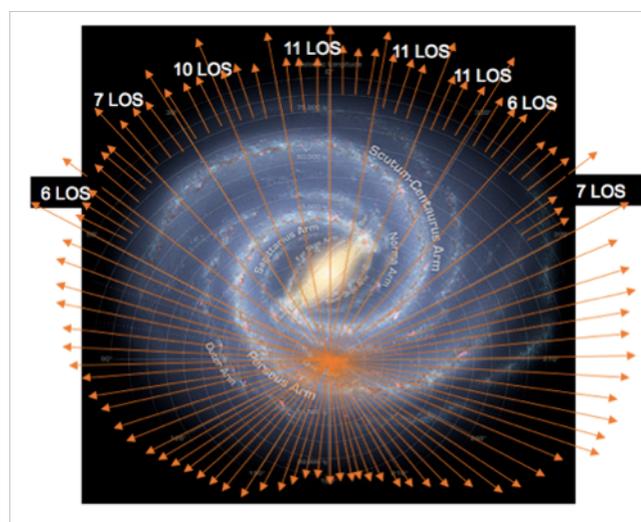

**Figure 2:** GOTC+ observations along line of sights resulted in an under-sampled survey (from Pineda et al 2010)

The Herschel Space Observatory highlighted the immense potential of far-infrared spectroscopy in understanding the complex physics of the ISM. While the superb imaging capabilities of Herschel resulted in panoramic views of the Galactic Plane that provided a full census of the dust reservoir available, there is a distinct lack of similar information for the gaseous component of the ISM. Only then, will be able to establish the gas-to-dust ratio and how this affects star formation in our own Galaxy and in external galaxies. But Herschel spectroscopic observations that allow us to probe the gas phase structure of the ISM were limited to less than 1% of the whole sky. Our mission concept, the Far-Infrared Spectroscopic Explorer (FIRSPEX) comes to fill this gap: FIRSPEX will carry out a large area survey in four discreet spectral channels centered on key FIR lines: C+ 158, N+ 205, C and CO(6-5). The choice of the lines has been motivated by the need to study the atomic, ionized and molecular clouds in the ISM. The FIRSPEX mission concept presented corresponds to an ESA Small Mission configuration. However, expanding FIRSPEX (as FIRSPEX++) to a larger (M-class) mission configuration would enable a much more capable mission.

## 3.     FIRSPEX Mission Goals

The top priority of the FIRSPEX mission is to carry out a wide area high spectral resolution survey of the phase structure of the ISM in our own Galaxy, focusing primarily on the Galactic Plane (the FIRSPEX Enhanced Sky Survey, ESS). This will be achieved by mapping, in high spectral resolution, the spectral tracers sensitive to the various phases of the ISM. While wide-area dust continuum surveys with Herschel have been used to trace the dust mass of the ISM, the proposed FIRSPEX surveys will provide a census of the mass of the atomic, ionised and molecular gas. The four tracers have been carefully selected to probe the ionised, atomic and molecular ISM: **C+ 158µm (1.91 THz), N+ 205µm (1.46 THz), CI370 µm [890 GHz) and CO(6-5) 433µm (691 GHz).**

The FISPEX mission goals are:
1. Delineate the transition from atomic to ionized to molecular gas in our Galaxy and trace the amount of gas in each phase. To achieve this goal we will obtain fully sampled velocity-resolved observations of C+,N+,C and CO.
2. Investigate the impact of massive stars on the surrounding ISM
3. Trace the CO-dark H2 gas and determine the total mass of the ISM.
4. Improve the accuracy of the CIB spectrum
5. Explore the gas-phases of the ISM in nearby galaxies and study how these affect star formation
6. Establish templates that can be used to interpret ALMA C+ detection of high-z galaxies.

The gas in the ISM in our Galaxy and in external galaxies can be found in three distinct phases, atomic, ionized and molecular. Studying the common boundaries (interfaces) between these three phases is, however, more challenging. FIRSPEX will carry out surveys in the C+ 158 µm, N+ 205 µm, C 370 µm and CO(6-5) 433 µm transitions. These lines are key diagnostics of the interstellar medium: The C+ line dominates the cooling of diffuse atomic clouds, the so-called Cold Neutral Medium (CNM) as well as the surfaces of molecular clouds and is a key diagnostic of photo-dissociation regions (PDRs) where far-ultraviolet photons from massive stars impinge on the nearby surfaces of molecular clouds, heating, photo-dissociating and -ionizing the gas. The N+ line probes the Warm Ionized Medium (WIM). This 1 kpc thick layer is powered by on-going star formation. The C emission line arises primarily in interface regions between zones emitting in C+ and CO. As the ionisation potential of neutral carbon is quite close to the dissociation energy of CO, neutral carbon subsequently may be ionized rather easily. In principle observations of emission from CO, CI and CII provide significant information on the physical condition of the cloud complexes from which the emission arises. The cross-analysis of complete Galactic Plane mapping in C+ and N+, with the FIR dust maps provided by the Herschel Hi-GAL survey (Molinari et al. 2010) will enable us to trace the budget of C+ emission potentially arising in shocks as a function of the position in the Galaxy (e.g. Figure 3) allowing us to quantify the relative importance of gravitationally induced versus shock-dominated filament formation mechanisms in different regions of the Milky Way, from the Central molecular Zone to the outskirts of the Galaxy beyond the Solar circle.

In order to explore the properties of the gas-phase ISM of external galaxies and compare to those found in our own Galaxy we propose to target a sample of nearby galaxies. These observations will be carried out in pointed mode. The most important species for 'decomposing' the gas content of a galaxy are: N+ which can be used to estimate the ionized gas contribution of HII regions (e.g. Vasta et al. 2010, Braine et al. 2012, Rigopoulou et al. 2013, Oberst et al. 2011); C+ which is the best tracer of the PDR regions; C which is often correlated with $H_2$ and CO (e.g. Offner et al. 2014) and finally CO which traces the amount and distribution of molecular gas at large scale. The carbon-bearing species are the only ones that can trace the three phases while being also abundant enough to be easily detectable. In recent years a lot of effort has been invested in modelling C, C+ and CO observations from ISO, Herschel, as well as ground based telescopes such as IRAM and JCMT in nearby galaxies using PDR models (e.g. Rigopoulou et al 2013, Rosenberg et al. 2014, Pellegrini et al. 2013, Bisbas et al. 2014). A high spectral resolution map of nearby galaxies in CI, CII, and CO would yield the determination of their 3D structure which, when compared to photo-ionisation and PDR models (e.g. Figure 5)

High spectral resolution mapping of the C+ emission in our own and nearby galaxies will allow us to establish the validity of C+ as a universal tracer of Star Formation activity in the high redshift Universe. One of the main drivers of ALMA is the ability to detect C+ in normal galaxies in the high redshift Universe. FIRSPEX observations will be fundamental to transforming these ALMA measurements into quantitative science.

The proposed FIRSPEX survey can also be used to remove the uncertainty on the current measurement of the Cosmic Infrared Background (CIB) integrated flux thus, improving upon the original COBE measurements. The current generation of far-IR and submm observatories although capable of detecting individual sources contributing to the CIB, are incapable of improving on the absolute flux measurements of the CIB obtained by COBE. FIRSPEX heterodyne receivers will take a different approach. Their narrow frequency bands are not sensitive enough to detect individual CIB sources (though this may be possible for bright individual sources in the deepest survey regions around the ecliptic poles), but they can accurately measure the total absolute flux. FIRSPEX ESS allows weak signals across many pixels to be combined to produce an integrated detection of any diffuse integrated background signal, such as the CIB. We intend to use the ``broad-band'' intensities measured in each of the FIRSPEX channels to measure the integrated flux of the CIB to much greater precision than was possible from COBE. In this way we will remove the current uncertainty on the integrated flux in the CIB and allow emission from the dust obscured universe to be better understood. The best measurement of the CIB SED will be provided by combining the signals received by FIRSPEX over a large area of sky and at the four channels, similar to the approach applied by Puget et al. (1996). In Figure 4 we show the sensitivity that would be reached by our observations over a 10° x 10° patch of sky in the ``all sky mode'' of the FIRSPEX for the 691 GHz channel. An absolute measurement of the flux of the CIB is of increasing importance as we delve deeper into the sources that produce it. However, it is currently known to precisions no better than 30\%. FIRSPEX has considerable potential for improving on the observations made with COBE in this area, though the requirements for such a project have implications on instrument and survey design and execution.

Finally, FIRSPEX will have the ability to detect C+ emission from intermediate redshift galaxies (in a narrow redshift band) in a redshift where the Star Formation Redshift Density (SFRD) of the Universe is increasing rapidly (e.g. Magnelli et al. 2011). With FIRSPEX we will be able to detect [CII] emission from (z~0.3) as well as (z~1.3) ULIRGs, covering important epochs in the star formation activity of the Universe investigating the important $L_{CII}$–$L_{IR}$ relation (see Figure 6). We stress that C+ observations of galaxies in both of these redshift ranges are currently inaccessible from the Earth and will complement observations of C+ from high redshift galaxies with ALMA.

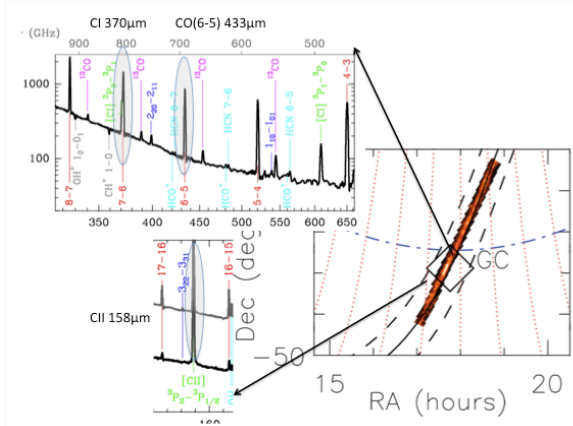
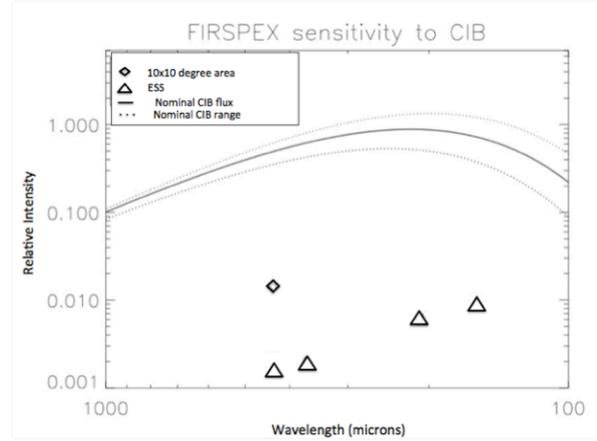

**Figure 3:** The Hi-GAL survey will be part of the FIRPSEX ESS (the ESS strip is denoted by the dashed line). The Spectra on the left are taken with PACS and SPIRE towards Sgr A but at much lower spectral resolution than that afforded by FIRSPEX (Goicoechea et al.2013)

**Figure 4:** The range of CIB spectral energy distributions allowed by current observations and the ability of FIRSPEX to set tight new constraints on these measurements.

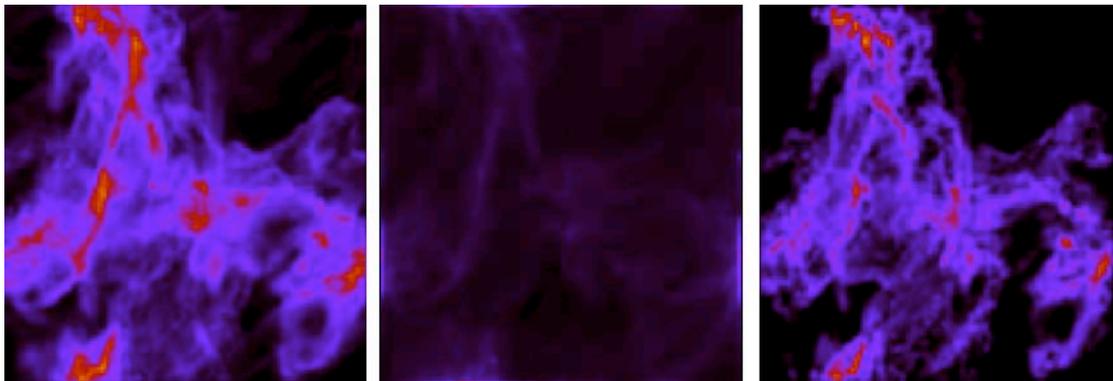

**Figure 5:** 3-dimensional synthetic maps (computed with the ORION hydrodynamical and the 3DPDR codes) showing neutral carbon C (left), ionized carbon C+ (middle) and CO(1-0) maps (right).

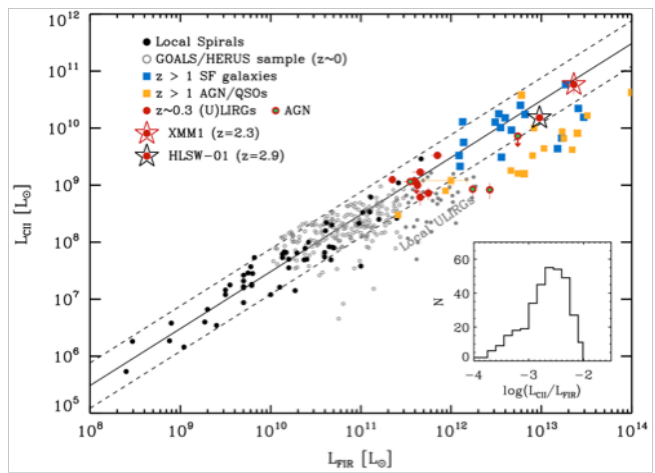

**Figure 6:** C+ emission line luminosity as a function of far-IR luminosity, including the majority of galaxies with available [CII] observation up to date (from Magdis et al. 2014).

## 4. FIRSPEX Orbit and Payload
### 4.1 Orbit

The chosen Low Earth Orbit (LEO) corresponds to a 96 min orbit and a 6 months period to cover the entire sky. We propose the Enhanced Sky Survey centered on a wide strip in the Galactic Plane and two areas centered on the North/South Ecliptic Pole as shown in Figure 7. The FIRSPEX Enhanced Sky Survey (ESS) will focus on two areas, one in the Galactic Plane and one in each of the Ecliptic Poles covering a total of 5,000 sq. deg. in the Sky. ESS will map the large scale emission of C+, N+, C and CO(6-5) in a +/- 5° wide strip above and below the Galactic Plane (GP) and an additional cap in the Ecliptic Poles (EPs, from +/-70° to +/-90°). This will be the first time that it will be possible to probe the line emission at high galactic latitudes with spectrally resolved observations. In addition FIRSPEX will carry out an all-sky survey in `direct imaging mode' at the using the lowest frequency (691 GHz) channel only.

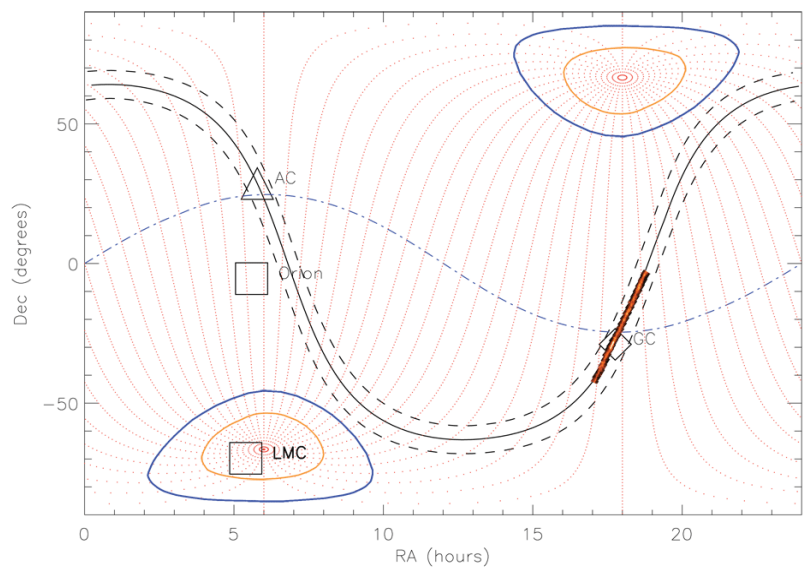

**Figure 7:** ESS Coverage: black dashed lines denote the area (+/- 5°) around the Galactic Plane; the position of the GC is clearly marked. Overlaid is the HiGAL survey (Molinari et al. 2010). The contours around the N&S EPs denote the proposed area to be covered: blue contour (+/-70° to +/-90°) is the goal coverage and yellow contour (+/-78° to +/-90°) is the minimum coverage (the exact area to be determined during Phase A).

### 4.2 Payload

The FIRSPEX payload (in current S-class configuration, see Table 1, Figure 9) comprises four heterodyne detection bands targeting key molecular and atomic transitions in the THz and supra-THz (>1THz) frequency range. The band frequencies of 2 x 1.9 THx, 2x 1.46 THz, 1x 890 GHz, 1x 691 GHz correspond to the [CII], [NII], [CI] and CO(6-5) lines respectively (see Figure 8). Although Superconducting receivers operating at ~4K temperature represent the most sensitive technology for heterodyne systems operating in space, present-day space cooler technology cannot achieve such a low temperature with the available heat rejection and power generation capabilities from a small satellite like FIRSPEX placed in Low Earth Orbit. Fortunately, semiconductor detectors do not have such stringent cooling requirements and, although less sensitive than superconducting detectors, offer an attractive alternative and are capable of meeting the scientific performance requirements of the FIRSPEX mission. FIRSPEX uses semiconductor Schottky diode mixers as the primary detection elements in each of the four high- spectral resolution heterodyne receivers. Each receiver will operate within a well-defined spectral range corresponding to the key species identified above. The receivers offer excellent signal-to-noise with double sideband system noise temperatures of ~950K at 0.69THz and ~6300K at 1.9THz with four wide-band digitizing spectrometers each with instantaneous bandwidth of 2 GHz and 1MHz spectral resolution. Adequate local oscillator (LO) power, a crucial element of the heterodyne mixing process, will be provided by sub-harmonic mixers using harmonic frequency up-conversion. The mixers and LOs can operate at 100 K, with cooling provided by a previously space-qualified and demonstrated Stirling cycle mechanical cooler. The spacecraft will carry a 0.85-m class sun-shielded telescope operating at ambient temperature and with three-axis stabilized pointing provided via the spacecraft attitude control system. The optical performance requirements on the telescope are relaxed compared to an optical telescope and we expect to use the same technology used for Herschel. The mission will operate in both survey and, for pointed observations, observatory mode. The lifetime is expected to be 2-3 years.

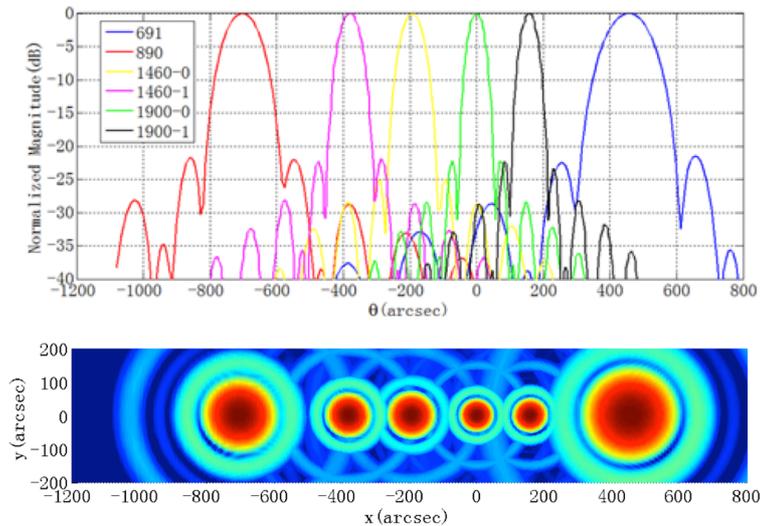

**Figure 8:** FIRSPEX instantaneous-FOV. Upper plot is normalised far-field response profile. Lower plot is instantaneous FOV as a contour plot in dB.

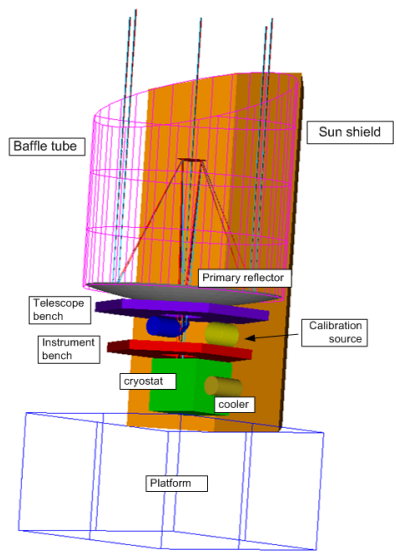

**Figure 9:** Outline model of the FIRSPEX payload showing the relationship between the various payload elements and a notional service module

**Table 1:** FIRSPEX Mission Overview

| | FIRSPEX – Mission Summary |
|---|---|
| Key Science Questions to be Addressed | • What is the origin and evolution of the Interstellar Medium in our Galaxy?<br>• What determines the lifecycle of the interstellar Medium in the Milky Way?<br>• What are the physical conditions of the gas clouds in our own Galaxy and in Nearby Galaxies?<br>• Is C+ a reliable tracer of Star Formation in the Universe? |
| Science Objectives | • Determine the physical properties of the three phases of the ISM and the relationship between them.<br>• Establish the impact of massive stars on their environment<br>• Trace the CO-dark H2 gas and determine the total mass of the ISM<br>• Explore the gas properties of the ISM in nearby galaxies and study how Star Formation is affected<br>• Establish templates to interpret C+, N+ and C detections of high-z galaxies with ALMA |
| Core Survey | • Enhanced Sky Survey: well-sampled survey of the Galactic Plane and the Ecliptic Poles (5000 sq. deg)<br>• All sky survey at 433 μm<br>• Pointed follow up observations of galactic targets<br>• Pointed observations of nearby galaxies<br>• Pointed observations of selected high-z and lensed galaxies, (C+, N+ detections) |
| Observational Strategy | • High spectral resolution, well sampled survey of the Galactic Plane and the Ecliptic Poles<br>• Pointed Observations of Nearby Galaxies |
| Telescope | • SiC 0.85 cm diameter on axis Cassegrain based on Airbus-DS technology |
| Payload Instrument | • Six band heterodyne receiver based on Schottky diode mixers cooled to ~100 K<br>• Band frequencies: 2 x 1.9 THx, 2x 1.46 THz, 1x 890 GHz, 1x 691 GHz<br>• Diffraction limited beams<br>• Sensitivities - Tsys DSB ~13000 K; 5000 K; 2060 K; 1500 K |
| Spacecraft | • S/C Dry mass ~300 kg<br>• Dimensions: ~1x1x2 m maximum. Based on Surrey Satellite Technology Limited CHEOPS definition study depending on payload configuration.<br>• Pointing requirements: pointed coarse APE of 36 arcsec; pointed RPE 12 arcsec; scan stability 0.125 arcsec/second; AKE 9 arcsec all 3σ values<br>• Attitude control system: reaction wheels and star trackers<br>• Thermal Control System<br>  - Passive cooling of telescope via ~250 K via telescope baffle<br>  - Miniature Stirling Cycle Cooler for active cooling to ~100K for THz receivers ~1W heat lift required<br>  - Dedicated radiator for cryostat intercept<br>• Telecommand, Telemetry and Communication: S-band, 2 GBit of science data per day transmitted to ground stations in Kiruna and UK. |
| Launcher, Orbit, Mission Phases and Operations | • Based on CHEOPS mission profile<br>• China launch, Long March 2D to 600 km SSO with 06:00 Local Time of Ascending Node Launch in 2020's.<br>• Nominal mission duration 2 years plus 6 months transit, cooldown and commissioning.<br>• MOC shared between China and ESA, Instrument Operations and Science Data Centre distributed between National agency supported institutes in Europe and China.<br>• One or more ground contact sessions per day for telecommand uplink and science downlink. |
| Data Policy | • Large maps of Galactic Plane and Ecliptic Poles released as soon as possible.<br>• AO for Open time - short proprietary period after nominal SNR is reached |

## 5. FIRSPEX Scientific Requirements

### 5.1 Sky Coverage

The FIRSPEX survey aims to cover a minimum of ~5,000 square degrees of sky in four discreet spectroscopic bands. We concentrate on two areas of the sky, the first is a 10 degree-wide strip (+/-5°) above and below the Galactic Plane (3,800 sq. deg). The second wide area comprises the North and South Ecliptic Poles (SEP, NEP) and extends from +/-70° to +/-90° covering a total area of 2500 sq. degrees (Figure 7). We call this survey the Enhanced Sky Survey (ESS). In addition to the ESS the entire sky will be covered in the 433μm band. The driver for the `direct detection' channel is to improve the precision of the CIB measurements and to deliver a `foreground dust ' map and does not impose any additional operational constraints. About 80% of sky is available to FIRSPEX for pointed observations. Target acquisition per orbit will be for 35 minutes approximately depending on location on the sky. The FIRSPEX targets for pointed observations are distributed throughout the sky. The targets will be revisited multiple times until the required sensitivity is reached

### 5.2 Pointing Accuracy

The pointing stability for the ESS survey (for the purpose of beam reconstruction) has been defined to be approximately a quarter of the C+ beam, that is ~12". The same pointing stability is also required for pointed observations.

### 5.3 Sensitivity

The sensitivity of the ESS is driven by the ability to detect the C+ line in the Galactic Plane. Figure 10 shows the distribution of the main beam temperatures from the Herschel GOTC+ project (from Goldsmith et al. 2012). The [CII] emission (shaded region) reaches down to Tmb~0.5 K. Our goal for the FIRSPEX ESS is to reach down to ~0.2 K to ensure that we adequately sample faint diffuse C+ emission.

The required radiometric sensitivity is given of both in terms of the heterodyne system noise-temperature $T_{sys}$ (i.e. of a black-body source that would give instantaneous RF signal level equal to the noise), and in terms of the minimum detectable delta-T of such a source, after integration over the observing time. The table below shows the calculation for $T_{sys}$, using the equation

$$T_{sys} = T_{optics} + \frac{T_m}{G_{optics}} + \frac{T_{LNA}}{G_{optics} G_m} + \cdots$$

and for delta-T using the equation

$$\frac{\Delta T}{T_{sys}} = \frac{1}{\sqrt{\Delta \nu \tau}}$$

Where $\Delta \int$ is the spectral resolution of 2 MHz, $|$ is integration time 4seconds

Assuming $T_m$ ~6310 K for the C+ receiver (Table 2) we estimate that we can reach 1.7K in a single survey pass in the 15′x0.8′ beam (for C+). But since we employ two C+ receivers we actually achieve 1.2 K per 15x16 arcmin spaxel and assuming we can co-add these we get a 15x15 arcmin `skybeam' (ie ~9 beams) we get 0.4 K. We propose to scan the sky 3 x therefore our final sensitivity is ~0.22 K which matches our sensitivity requirement.

The required sensitivities for the pointed observations of nearby galaxies have based on prior observations with ISO-LWS (Brauher et al 2008) and more recent ones with Herschel (Rigopoulou et al. 2013, Rosenberg et al. 2014, Tabatabei et al. 2013). The C+ line fluxes range from a few x10$^{-15}$ up to x10$^{-17}$ Wm$^{-2}$ which are well within the reach of FIRSPEX. The required integration times range from 1hr to 4hr (including those galaxies where more than one beams are required to sample the emission).

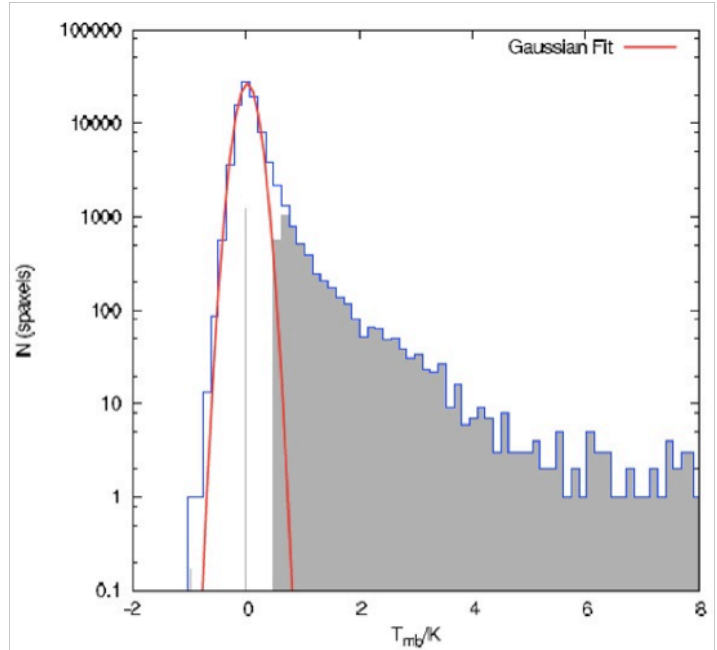

**Figure 10:** Histogram of main beam temperatures from the GOT-C+ survey. The obvious peak, centered around zero K, is due to the radiometric noise from HIFI. The distribution of the noise is quite well represented by the Gaussian shown. The [CII] emission, indicated by the shaded region, is the excess above this noise, and varies between 0.5 K <Tmb< 8 K.

**Table 2: FIRSPEX Channel Sensitivity Table**

| **Band frequency** | **691GHz** | **890GHz** | **1460GHz** | **1900GHz** |
|---|---|---|---|---|
| Scottky mixer f | 342.75 GHZ | 442.13GHz | 727.01GHz | 473.45GHz |
| $G_{optics}$ | 0.7 | 0.68 | 0.62 | 0.56 |
| $T_{optics}$ (K)* | 30 | 32 | 28 | 44 |
| $G_m$ | 8 | 9 | 12 | 18 |
| $T_m$ (100K) | 950 | 1260 | 2840 | 6310 |
| $G_{LNA}$ | 10000 | 10000 | 10000 | 10000 |
| $T_{LNA}$ (100K)** | 15 | 15 | 15 | 15 |
|  | … | … | … | … |
| **T$_{sys}$ (K)** | **1523** | **2060** | **5000** | **13000** |
| **ΔT (K)** | **0.54** | **0.73** | **1.77** | **4.60** |

\* The temperature of the optics is set to be 100K.
\*\*The $T_{LNA}$ estimated by assuming the noise of the LNA is linearly dependent to the temperature

## 5.4  Velocity Coverage

The expected velocity ranges for the ESS have been defined from Figure 11 (taken from Pineda et al. 2013). The FIRSPEX requirement has therefore been set to -150 km/sec to +150 km/sec for C+, N+ ,C and CO(6-5) channels.

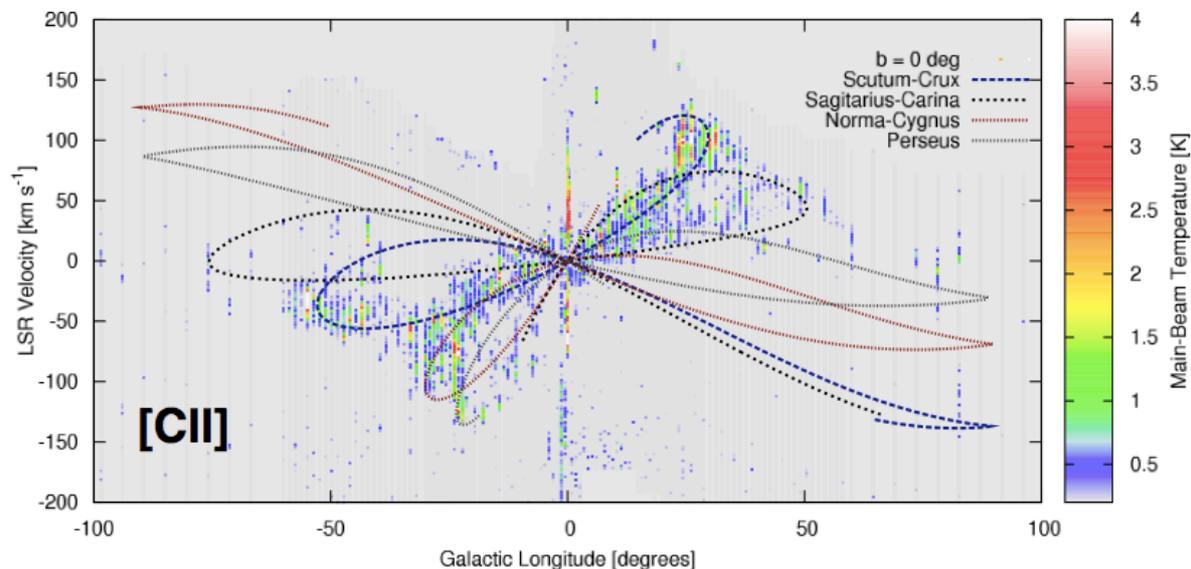

**Figure 11:** Position velocity map for C+ (between l = -100° and +100° at  b = 0°, from the GOTC+) survey. The FIRSPEX ESS velocity range requirement is set to [-150, +150] km/sec for the C+,N+, C and CO(6-5) observations.

There are two requirements for the expected velocity ranges for nearby galaxies. The first requirement refers to galaxies with wide velocity ranges (e.g. LMC) [-500  +500] km/sec where re-tuning of the LO will be required (described by the term `narrow tuning).

For galaxies with `narrower' emission line profiles the requirement is [-300, +300] km/sec, based on line widths for nearby galaxies (selected from e.g. Brauher et al. 2008). . In both cases the velocity range includes the relative shift due to the galaxy's redshift (cz).

## 5.5  Bandwidth Requirements

Table 3**Error! Reference source not found.** reports the required bandwidths to make the spectral measurements reported in Section 5.4

**Table 3:** Required bandwidth for spectral measurements

| **Species** | **CO (6-5)** | **CI[609]** | **NII[205]** | **CII[158]** |
|---|---|---|---|---|
| **Frequency (GHz)** | **691** | **809** | **1450** | **1900** |
| **Wavelength (μm)** | **433.9** | **370.6** | **206.8** | **157.8** |
| | **Frequency (GHz)** | **Frequency (GHz)** | **Frequency (GHz)** | **Frequency (GHz)** |
| ESS (Galactic Plane and EPs) | 0.7 | 0.8 | 1.5 | 1.9 |
| Local Galaxies (all) [1] | 2.3 | 2.7 | 4.8 | 6.3 |
| Local Galaxies [2] | 1.4 | 1.6 | 2.9 | 3.8 |
| High recession velocity galaxies | 4.6 | 5.4 | 9.6 | 12.7 |
| Direct Detection[3] | 10 | | | |

**1:** applies to all nearby galaxies including those whose lines are expected to be wide (ie 500 km/sec)
**2**: applies to nearby galaxies with line-widths ~300 km/sec

3: requirement for the Direct Detection channel for the All Sky survey

## 5.6 Spectral Resolution

The required resolution per channel is quoted in Table 4

**Table 4:** Required spectral resolution for target line species.

| Species | CO(6-5) | [CI] | [NII] | [CII] |
|---|---|---|---|---|
| **Frequency (GHz)** | 433.9 | 809 | 1460 | 1900 |
| **Resolution (MHz)** | 1.0 | 1.3 | 2.5 | 3 |

## 5.7 Mission Lifetime

Each orbit requires 96 mins and one full pass of the sky will require 6 months. To reach the required sensitivities and to provide some additional redundancy in the survey we propose to repeat the survey 3 times, hence the ESS required 18 months to completion. This is the absolute minimum mission duration. The CfP defines as mission lifetime 2 to 3 year. Assuming a 2 year mission we propose to devote 18 months to the survey and 6 months to pointed observations of nearby galaxies and/or follow up pointed observations on interesting targets discovered through the ESS. Assuming a 3 year mission we will complete the full sample of nearby and distant galaxies.

## 6. Summary

The FIRSPEX Mission provides a unique opportunity to extend the All-Sky maps provided by the previous IRAS and AKARI missions to the spectral domain. FIRSPEX will survey the galactic plane in the C+ 158 μm, N+ 205 μm, C 370 μm and CO (6-5) 433 μm. These lines are key diagnostics of the interstellar medium. The FIRSPEX data can be combined with a multitude of existing surveys, including line surveys – the 21cm line tracing HI in diffuse clouds, the 2.6mm CO J=1–0 transition tracing molecular clouds, & the Hα survey, tracing the WIM – and continuum surveys – the IR continuum tracing warm dust in regions of massive star formation (see Section 2.3.2.2) as well as cold dust in atomic and molecular clouds, radio emission at several wavelength tracing synchrotron emission due to relativistic electrons gyrating in the interstellar magnetic field, as well as free-free emission from gas ionized by massive stars, X-ray surveys of the tenuous hot gas (the Hot Intercloud Medium) generated by supernova explosions and filling the superbubbles and chimneys in the ISM, and γ-ray surveys tracing the interaction of cosmic rays with atomic and molecular gas. The proposed FIRSPEX survey – together with this ancillary data – will address key questions on the origin and evolution of the interstellar medium such as: the energetics of the CNM & the WIM, the kinematics of the interstellar medium, the evolutionary relationship of atomic and molecular gas, the relationship of these ISM phases with newly formed stars, and the conversion of their radiative and kinetic power into thermal and turbulent energy of the ISM. The proposed survey will thus provide key insight in the lifecycle of the interstellar medium of the Milky Way.

Finally, it is worth stressing that one of the main technical drivers of ALMA is the ability to detect the C+ 158 μm in normal galaxies in high redshift universe. Early results are already routinely used to determine the star formation rate in the early universe, despite that its origin in the Milky Way is not understood. FIRSPEX will be key to turn these observations of the early universe into quantitative science.

## References

Bisbas, TG, Bell, TA, Viti, S,  et al 2014, MNRAS, 433, 111
Brauher, JR, et al, 2008, ApJS, 178, 280
De Graauw, T, Helmich, F.P., Phillips, TG, et al., 2010, A&A, 518, 6
Fixsen, DJ,  Bennet, CL Mather, JC, 1999, ApJ 526, 207
Fixsen, DJ, Dwek, E, Mather, JC, et al, 1998, ApJ, 508, 123
Kirk, H., et al., 2010, ApJ, 723, 457
Langer, WD, Velusamy, T, Pineda, JR, et al. 2010, A&A, 521, 17
Magdis, GE, et al, 2012, ApJ, 760, 6
Magdis, GE, Rigopoulou, D, Hopwood, R, et al, 2014,ApJ 796, 63
Magnelli, B, Elbaz, D, Chary, RR, et al 2011, A7A 528, 35
Molinari, S, Swinyard, BM, Barlow, M, et al 2010, A&A, 518, 100
Nakagawa, T, Yui, Y, Yukari, Y. et al., 1998, ApJS, 115, 259
Oberst , TE, Parshley, SC, Stacey, GJ, et al. 2006, 652, 125
Offner, S, Bisbas, T, Bell, T, et al 2014, MNRAS, 440, 80
Pellegrini, E,W, Smith, JD, wolfire, MG, et al 2013, ApJ, 779, 19
Pilbratt, GL, Riedinger,  JR,  Passvogel, T, et al. 2010, A&A 518, 1
Pineda, JR,  Velusamy, T, Langer, WD, et al. 2010, A&A 521, 19
Puget, J-L, Abergel, A, Bernard, J-P, et al. 1996, A&A 308, 5
Rigopoulou, D, Hurley, PD, Swinyard, BM, et al 2013, MNRAS, 434, 2051
Rosenberg, MJF,  et al, 2014, A&A, 564, 126
Vasta, M, Barlow, M, Viti, S, et al 2010, MNRAS, 409, 1910
Velusamy, T, Langer, WD, Pineda, JR et al. 2012, A&A 541, 10